# Extra Gain: Improved Sparse Channel Estimation Using Reweighted $\ell_1$-norm Penalized LMS/F Algorithm

Guan Gui and Li Xu

Dept. Electronics and Information Systems
Akita Prefectural University
Akita, Japan
{guiguan, xuli}@akita-pu.ac.jp

Fumiyuki Adachi

Dept. Communications Engineering, Graduate
School of Engineering, Tohoku University
Sendai, Japan
adachi@ecei.tohoku.ac.jp

*Abstract*—The channel estimation is one of important techniques to ensure reliable broadband signal transmission. Broadband channels are often modeled as a sparse channel. Comparing with traditional dense-assumption based linear channel estimation methods, e.g., least mean square/fourth (LMS/F) algorithm, exploiting sparse structure information can get extra performance gain. By introducing $\ell_1$-norm penalty, two sparse LMS/F algorithms, (zero-attracting LMSF, ZA-LMS/F and reweighted ZA-LMSF, RZA-LMSF), have been proposed [1]. Motivated by existing reweighted $\ell_1$-norm (RL1) sparse algorithm in compressive sensing [2], we propose an improved channel estimation method using RL1 sparse penalized LMS/F (RL1-LMS/F) algorithm to exploit more efficient sparse structure information. First, updating equation of RL1-LMS/F is derived. Second, we compare their sparse penalize strength via figure example. Finally, computer simulation results are given to validate the superiority of proposed method over than conventional two methods.

*Keywords—Adaptive sparse channel estimation; zero-attracting least mean square/fourth (ZA-LMS/F); reweighted $\ell_1$-norm sparse penalty; compressive sensing.*

## I. INTRODUCTION

Recently, the wireless broadband transmission is becoming more and more important [3], [4]. The broadband signal is significantly distorted by frequency-selective fading and hence, some powerful equalization techniques need to be adopted. Any equalization technique requires accurate channel state information. Based on the assumption of dense finite impulse response (FIR), traditional linear channel estimation methods, e.g. standard least mean square/fourth (LMS/F) algorithm [5] have been proposed. However, FIR of real channel is often modeled as sparse and many of channel coefficients are zero. Because the relative longer discrete channel is sampled with higher sampling frequency (due to broader baseband transmission) according to the Nyquist-Shannon sampling theory while the significant channel coefficients are very few [6]. For a better understanding the concept of sparse channel, we give a figure example to introduce intuitively the relationship between number of channel taps and sampling frequency (bandwidth) in Fig. 1. Considering any *N*-length channel vector $w = [w_0, w_1, \cdots, w_{N-1}]^T$, sparseness of channel vector $w$ can be measured [7] by

$$\xi(w) = \frac{N}{N - \sqrt{N}} \left( 1 - \frac{\|w\|_1}{\sqrt{N}\|w\|_2} \right), \quad (1)$$

where $\|w\|_1$ and $\|w\|_2$ stands for $\ell_1$ norm and $\ell_2$ norm of $w$, respectively, i.e., $\|w\|_1 = \sum_i |w_i|$ and $\|w\|_2 = \sqrt{\sum_i w_i^2}$. Larger value of $\xi(w)$ implies sparser channel and vice versa. For example, sparseness of $\xi(w_d) = 1$ for $w_d = [1, 0, ..., 0]^T$ while $\xi(w_u) = 0$ for $w_u = [1, 1, ..., 1]^T$. According to above sparseness measure function in (1), we simple classify either dense or sparse structure channels also classified in Tab. I.

TAB. I. CHANNEL STRUTURES WITH RESPECT TO DIFFERENT BANDWIDTH.

| Transmission bandwidth | 5MHz | 10MHz | 50MHz | 100MHz |
|---|---|---|---|---|
| Channel delay spread | $0.5\mu s$ | | | |
| No. channel taps | 5 | 10 | 50 | 100 |
| No. nonzero channel taps | 5 | 5 | 5 | 5 |
| Channel distribution | uniform | | | |
| Channel sparseness | 0 | 0.4283 | 0.7964 | 0.8427 |
| Channel structure | dense | Quasi-sparse | sparse | sparse |

To deal with the sparse problems, adaptive sparse channel estimation methods have been proposed to achieve performance gain by exploiting inherent channel sparse structure information [1], [8]–[11]. In these state-of-the-art methods, zero-attracting (ZA-LMS/F) and reweighted ZA-LMS/F (RZA-LMS/F) [1] have been validated as one of effective methods with $\ell_1$ norm sparse constraint. Either zero-attracting (ZA) or reweighted ZA (RZA), sparse constraint ability is limited due to the fact that $\ell_1$ norm sparse solution. It is well known that $\ell_1$ norm solution is only a suboptimal solution where exists an obvious performance gap to the optimal solution [12]. Please notice that finding the optimal solution is a NP hard problem [12]. Hence, more effective sparse approximation can reduce the performance gap. In [2],

E. Candes proposed an improved sparse solution by using reweighted $\ell_1$ norm (RL1) sparse function. Motivated by this work, we propose an improved sparse LMS/F algorithm by introducing the RL1 to standard LMS/F. First of all, we derive the updating equation of RL1-LMS/F algorithm and the sparse penalty strength of different sparse functions is compared as well. Later, computer simulation examples are given to verify our propose method.

Section II introduces system model and reviews conventional sparse LMS/F algorithms. In Section III, an improved sparse LMS/F based channel estimation methods is proposed. In section V, computer simulation results are given and their performance comparisons are also discussed. Concluding remarks are resented in Section V.

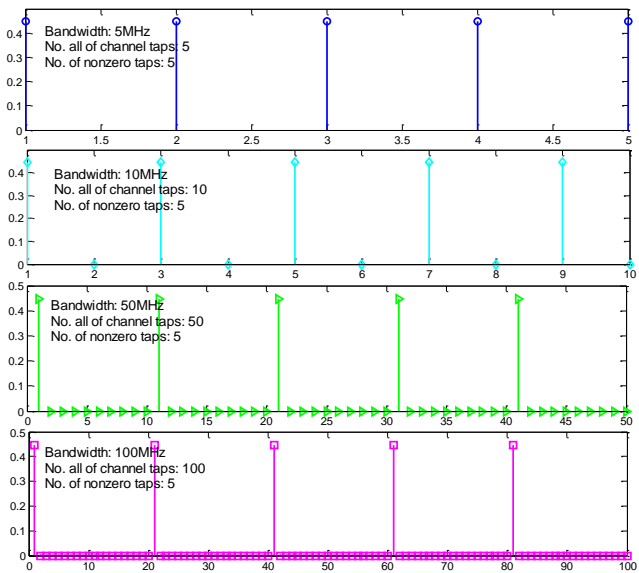

Fig. 1. Number of channel taps relates to baseband transmission bandwidth.

## II. OVERVIEW OF CONVENTIONAL SPARSE LMS/F ALGORITHMS

Considering a FIR based broadband wireless communication system, the input signal vector $\boldsymbol{x}(k)$ and output signal $d(k)$ are related by

$$d(k) = \boldsymbol{w}^T \boldsymbol{x}(k) + z(k), \qquad (2)$$

where $\boldsymbol{w} = [w_0, w_1, ..., w_{N-1}]^T$ is a $N$-length unknown FIR channel vector which is supported only by $K$ significant coefficients ($K \ll N$); $\boldsymbol{x}(k)$ is a $N$-length input signal vector $\boldsymbol{x}(k) = [x(k), x(k-1), ..., x(k-N+1)]^T$ and $z(k)$ is an additive Gaussian noise variable satisfying $\mathcal{CN}(0, \sigma_n^2)$. The object of adaptive sparse channel estimation is to probe the unknown FIR channel vector $\boldsymbol{w}$ with $\boldsymbol{x}(k)$ and $d(k)$. According to Eq. (2), channel estimation error $e(n)$ is:

$$e(k) = d(k) - \tilde{\boldsymbol{w}}^T(k) \boldsymbol{x}(k), \qquad (3)$$

where $\tilde{\boldsymbol{w}}(k)$ is assumed as ZA-LMS/F channel estimate. Based on Eq. (3), cost function of ZA-LMS/F [1] can be written as

$$G_{ZA}(k) = \underbrace{\frac{1}{2} e^2(k) - \frac{1}{2} \phi \ln\left(e^2(k) + \phi\right)}_{Estimation\ error} + \lambda_{ZA} \underbrace{\|\tilde{\boldsymbol{w}}(k)\|_1}_{Sparsity}. \qquad (4)$$

where $\phi$ is an user setting positive parameter; $\lambda_{ZA}$ is a regularization parameter to balance the estimation error and sparsity of $\|\tilde{\boldsymbol{w}}(k)\|_1$; sgn($\cdot$) is the sign function which operates on every component of the vector separately and it is zero for $x = 0$, 1 for $x > 0$ and $-1$ for $x < 0$. Hence, the update equation of ZA-LMS/F adaptive channel estimation is derived by

$$\begin{aligned}\tilde{\boldsymbol{w}}(k+1) &= \tilde{\boldsymbol{w}}(k) + \mu \frac{\partial G_{ZA}(k)}{\partial \tilde{\boldsymbol{w}}(k)} \\ &= \underbrace{\tilde{\boldsymbol{w}}(k) + \frac{\mu e^3(k) \boldsymbol{x}(k)}{e^2(k) + \phi}}_{LMS/F} - \rho_{ZA} \operatorname{sgn}(\tilde{\boldsymbol{w}}(k)),\end{aligned} \qquad (5)$$

$$\underbrace{\phantom{}}_{ZA-LMS/F}$$

where $\rho_{ZA} = \mu \lambda_{ZA}$; $\mu \in (0, 2/\gamma_{\max})$ is a step size of gradient descend step-size and $\gamma_{\max}$ is the maximum eigenvalue of the covariance matrix of $\boldsymbol{x}(k)$. The cost function of RZA-LMS/F is written as

$$G_{RZA}(k) = \underbrace{\frac{1}{2} e^2(k) - \frac{1}{2} \phi \ln\left(e^2(k) + \phi\right)}_{Estimation\ error} + \lambda_{RZA} \underbrace{\sum_{i=0}^{N-1} \log\left(1 + |\tilde{w}_i|/\varepsilon\right)}_{Sparsity}, \qquad (6)$$

where $\lambda_{RZA} > 0$ is *a regularization parameter* as well to balance the estimation error and sparsity of $\sum_{i=0}^{N-1} \log(1 + |\tilde{w}_i|/\varepsilon)$. The corresponding update equation is

$$\begin{aligned}\tilde{\boldsymbol{w}}(k+1) &= \tilde{\boldsymbol{w}}(k) + \mu \frac{\partial G_{RZA}(k)}{\partial \tilde{\boldsymbol{w}}(k)} \\ &= \underbrace{\boldsymbol{h}(k) + \frac{\mu e^3(k) \boldsymbol{x}(k)}{e^2(k) + \phi}}_{LMS/F} - \rho_{RZA} \frac{\operatorname{sgn}(\tilde{\boldsymbol{w}}(k))}{1 + \varepsilon |\tilde{\boldsymbol{w}}(k)|},\end{aligned} \qquad (7)$$

$$\underbrace{\phantom{}}_{RZA-LMS/F}$$

where $\rho_{RZA} = \mu \lambda_{RZA}/\varepsilon$ is a parameter which depends on step-size $\mu$, regularization parameter $\lambda_{RZA}$ and threshold $\varepsilon$. Here, reweighted factor is set as $\varepsilon = 20$ [1] to exploit channel sparsity efficiently. In the second term of (7), please notice that estimated channel coefficients $|\tilde{w}_i|$, $i = 0, 1, \cdots, N-1$ are replaced by zeroes in high probability if under the hard threshold $1/\varepsilon$. Hence, one can find that RZA-LMS/F can exploit sparsity and mitigate noise interference simultaneously.

## III. IMPROVED ADAPTIVE SPARSE CHANNEL ESTIMATION USING RL1-LMS/F ALGORITHM

The reweighted $\ell_1$ norm minimization for sparse channel estimation has a better performance than the standard $\ell_1$ norm minimization that is usually employed in compressive sensing [2]. It is due to the fact that a properly reweighted $\ell_1$ norm approximates the $\ell_0$ norm, which actually needs to be minimized, better than $\ell_1$ norm. Therefore, one approach to enforce the sparsity of the solution for the sparsity-aware LMS/F-type algorithms is to introduce the reweighted $\ell_1$ norm penalty term in thee cost function. Our reweighted $\ell_1$ norm penalized LMS/F algorithm considers a penalty term proportional to the reweighted $\ell_1$ norm of the coefficient vector. The corresponding cost function can be written as

$$G_{RL1}(k) = \underbrace{\frac{1}{2}e^2(k) - \frac{1}{2}\phi\ln\left(e^2(k)+\phi\right)}_{\text{Estimation error}} + \lambda_{RL1}\underbrace{\|f(k)\tilde{w}(k)\|_1}_{\text{Sparsity}}, \quad (8)$$

where $\lambda_{RL1}$ is the weight associated with the penalty term and elements of the $1 \times N$ row vector $w(k)$ are set to

$$[f(k)]_i = \frac{1}{\delta + |[\tilde{w}(k-1)]_i|}, \quad i = 0,1,\cdots,N-1, \quad (9)$$

where $\delta$ being some positive number and hence $[f(k)]_i > 0$ for $i = 0,1,...,N-1$. The update equation can be derived by differentiating (8) with respect to the FIR channel vector $\tilde{w}(k)$. Then, the resulting update equation is:

$$\begin{aligned}\tilde{w}(k+1) &= \tilde{w}(k) + \mu\frac{\partial G_{RL1}(k)}{\partial \tilde{w}(k)} \\ &= \tilde{w}(k) + \frac{\mu e^3(k)x(k)}{e^2(k)+\phi} - \mu\lambda_{RL1}\operatorname{sgn}\left(f(k)\tilde{w}(k)\right)\tilde{w}(k) \quad (10) \\ &= \underbrace{\tilde{w}(k) + \frac{\mu e^3(k)x(k)}{e^2(k)+\phi}}_{LMS/F} - \underbrace{\frac{\rho_{RL1}\operatorname{sgn}(\tilde{w}(k))}{\delta + |\tilde{w}(k-1)|}}_{RL1-LMS/F},\end{aligned}$$

where $\rho_{RL1} = \mu\lambda_{RL1}$. In Eq. (10), since $\operatorname{sgn}(f(k)) = \mathbf{1}_{1\times N}$, one can get $\operatorname{sgn}(f(k)\tilde{w}(k)) = \operatorname{sgn}(\tilde{w}(k))$. Note that although the weight vector $\tilde{w}(k)$ changes in every stage of this sparsity-aware LMS/F algorithm, it does not depend on $\tilde{w}(k)$, and the cost function $G_{RL1}(k)$ is convex. Therefore, the RL1 penalized LMS/F is guaranteed to converge to the global minimization under some conditions. To evaluate the sparse penalty strength of ZA, RZA and RL1, we define above three sparse penalty functions as follows:

$$\zeta_{ZA} = \operatorname{sgn}(w), \quad (11)$$

$$\zeta_{RZA} = \frac{\operatorname{sgn}(w)}{1+\varepsilon|w|}, \quad (12)$$

$$\zeta_{RL1} = \frac{\operatorname{sgn}(w)}{\delta+|w|}, \quad (13)$$

where channel coefficients in $w$ are assumed in range $[-1,1]$. Considering above sparsity functions in Eqs. (11)~(13), their sparse penalty strength curves are depicted in Fig. 2. One can find that ZA utilizes uniform sparse penalty to all channel coefficients in the range of $[-1,1]$ and hence it is not efficient to exploit channel sparsity. Unlike the ZA (11), both RZA (12) and RL1 (13) make use of adaptively sparse penalty on different channel coefficients, i.e., stronger sparse penalty on zero/approximate zero coefficients and weaker sparse penalty on significant coefficients. In addition, one can also find that RL1 (13) utilizes stronger sparse penalty than RZA (12) as shown in Fig. 2. Hence, RL1-LMS/F can exploit more sparse information than both ZA-LMS/F and RZA-LMS/F on adaptive sparse channel estimation. By virtue of Monte-Carlo based computer simulation, our proposed method will be verified in the following.

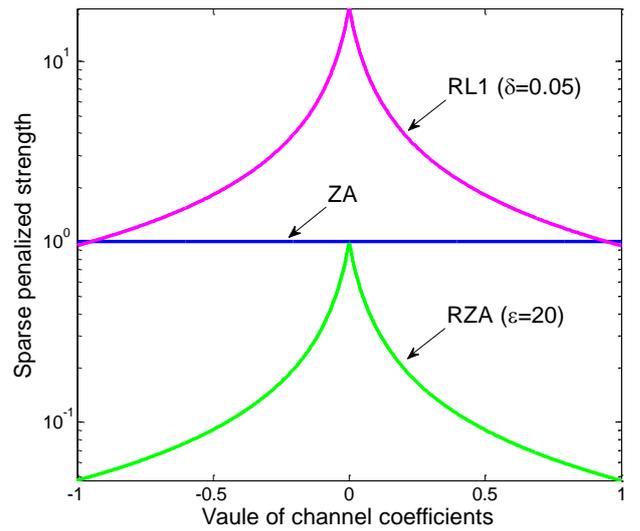

Fig. 2. Comparison of the three sparse penalty functions.

## IV. COMPUTER SIMULATION

Simulations are carried out to investigate the average MSE performance of the proposed ASCE methods using (R)ZA-LMS/F and RL1-LMS/F algorithms. The results are averaged over 1000 independent Monte-Carlo runs. The simulation setup is configured according to typical broadband wireless communication system in Japan [13]. The signal bandwidth is $60MHz$ located at the central radio frequency of $2.1GHz$. The maximum delay spread of $1.06\mu s$. Hence, the maximum length of channel vector $w$ is $N = 128$ and its number of dominant taps is set to $K \in \{4,8,16\}$. Dominant channel taps follow random Gaussian distribution as $\mathcal{CN}(0,\sigma_w^2)$ which is subject to $E\{\|w\|_2^2\} = 1$ and their positions are randomly decided within $w$. The received signal-to-noise ratio (SNR) is defined as $10\log(E_s/\sigma_n^2)$, where $E_s = 1$ is the unit transmission power. Here, we set the SNR as 10dB. All of the simulation parameters are listed in Table. II.

TAB. II. SIMULATION PARAMETERS.

| Parameters | Values |
|---|---|
| Training signal | Pseudo-random sequence |
| Channel length | $N=128$ |
| No. of nonzero coefficients | $K \in \{4, 8, 12\}$ |
| Distribution of nonzero coefficient | Random Gaussian $\mathcal{CN}(0, \sigma_w^2)$ |
| Threshold parameter for LMS/F-type | $\phi = 0.8$ |
| SNR | 10dB |
| Step-size | $\mu = 0.005$ |
| Regularization parameter | $\lambda_{ZA} = 0.00004$, $\lambda_{RZA} = 0.004$ and $\lambda_{RL1} = 0.00004$ |
| Parameter $\delta$ for RL1-LMS/F | $\delta = 0.05$ |
| Re-weighted factor for RZA-LMS/F | $\varepsilon = 20$ |

**Example 1**: Estimation performance of RL1-LMS/F verses threshold parameter $\phi$. Let us revisit Eq. (10) and rewritten it as RL1-LMS-like update equation:

$$\tilde{w}(k+1) = \tilde{w}(k) + \frac{\mu e^3(k) x(k)}{e^2(k) + \phi} - \frac{\rho_{RL1} \operatorname{sgn}(\tilde{w}(k))}{\delta + |\tilde{w}(k-1)|} \quad (14)$$
$$= \tilde{w}(k) + \mu(k) e(k) x(k) - \frac{\rho_{RL1} \operatorname{sgn}(\tilde{w}(k))}{\delta + |\tilde{w}(k-1)|},$$

where $\mu(k)$ is an variable-step-size (VSS):

$$\mu(k) = \frac{\mu e^2(k)}{e^2(k) + \phi}. \quad (15)$$

It is well known that the step-size is a critical parameter to balance the convergence speed and steady-state performance. In (15), $\mu(k)$ depends on two factors: $k$-th updating error $e(k)$ and threshold parameter $\phi$. Hence, setting the threshold parameter $\phi$ is very important for controlling the $\mu(k)$. We depict performance curves with different threshold parameters $\phi \in \{0.2, 0.4, 0.6, 0.8, 1.0\}$ in Fig. 3. One can find that larger $\phi$ brings more performance gain but meanwhile it incurs slower convergence speed. Hence, it is better to find the suitable $\phi$ so that $\mu(k)$ can balance the estimation performance and convergence speed. In the example 2, $\phi = 0.8$ is selected for compromising the stable steady-state performance but without scarifying much convergence speed.

**Example 2**: Estimation performance is evaluated in different number of nonzero channel coefficients, $K \in \{4, 8, 12\}$. To confirm the effectiveness of the proposed method, we compare them with sparse LMS/F algorithms, i.e., ZA-LMS/F and RZA-LMS/F. For a fair comparison, we utilize the same step-size $\mu$. In addition, to achieve approximate optimal sparse estimation performance, regularization parameters for two sparse LMS/F algorithms are adopted from the paper, i.e., $\lambda_{ZA} = 4 \times 10^{-5}$ for ZA-LMS/F; $\lambda_{RZA} = 4 \times 10^{-3}$ for RZA-LMS/F and $\lambda_{RL1} = 4 \times 10^{-5}$ for RL1-LMS/F. Average MSE performance comparison curves are depicted in Figs. 4~6. Obviously, RL1-LMS/F achieves better estimation performance than sparse LMS/F algorithms (ZA-LMS/F and RZA-LMS/F) in different channel sparsity. Indeed, RL1-LMS/F can obtain more performance gain in sparser channel such as $K = 4$ in Fig. 3. Figures clarify that the sparse LMS/F algorithms, i.e., ZA-LMS/F and RZA-LMS/F, achieve better estimation performance than LMS/F due to the fact that sparse LMS/F algorithms utilize $\ell_1$-norm sparse constraint function.

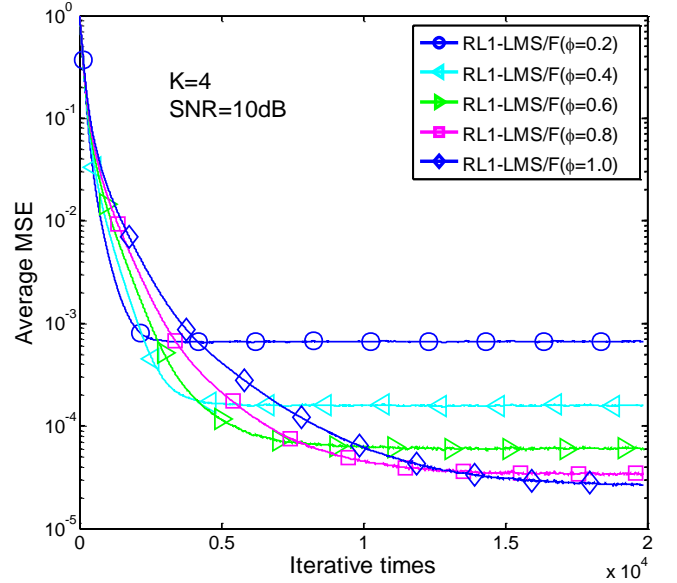

Fig. 3. Average MSE performance with respect to different threshold $\phi$.

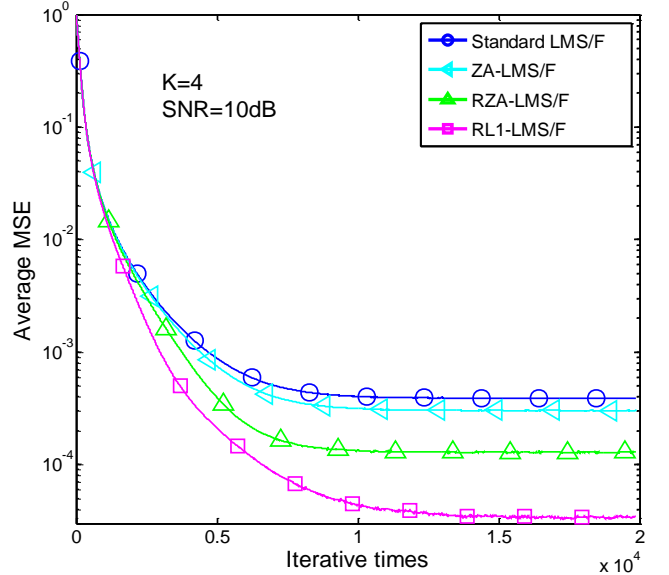

Fig. 4. Average MSE performance at $K = 4$.

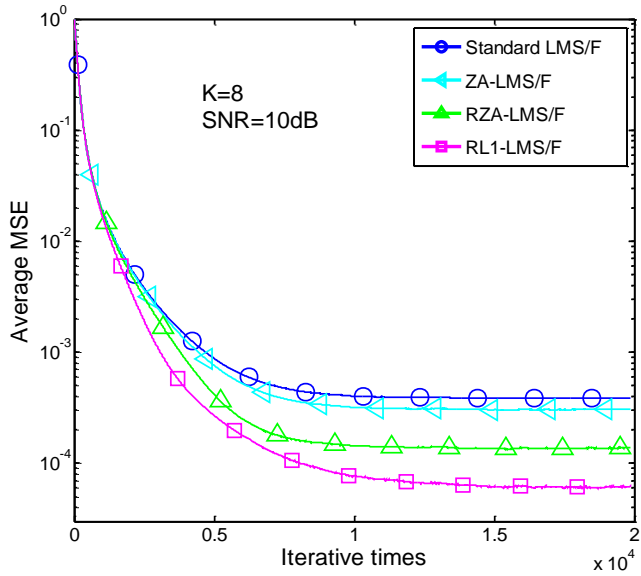

Fig. 5. Average MSE performance at $K=8$.

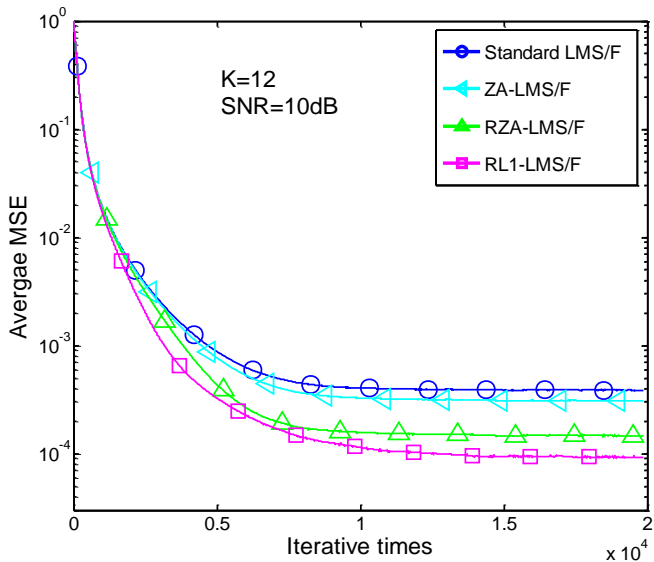

Fig. 6. Average MSE performance at $K=12$.

## V. CONCLUSION AND FUTURE WORK

In this paper, an improved RL1-LMS/F algorithm was proposed for estimating sparse channels in typical broadband wireless communications systems. We first revised two traditional adaptive sparse channel estimation methods, i.e., ZA-LMS/F and RZA-LMS/F. Inspired by re-weighted $\ell_1$-norm algorithm in CS, an improved adaptive sparse channel estimation method with RZA-LMS/F algorithm. For the better understanding of our motivation, the penalty ability in different sparse constraint functions was evaluated. In addition, by virtual of Monte Carlo simulation, we investigated threshold parameter approximate optimal regularization the selection of parameter for LMS/F-type algorithms. Based on the typical broadband wireless systems in Japan, simulation results showed that the proposed method achieves better estimation than traditional ZA-LMS/F and RZA-LMS/F.

One may notice that our proposed method depends on several parameters: regularization parameter $\lambda_{RL1}$, positive parameter $\delta$ as well as threshold parameter $\phi$. This paper only considered performance comparison with empirical parameters but lack of system analysis. In future work, we are going to give full performance analysis relates to these parameters.